\documentclass[%
 reprint,
 amsmath,amssymb,
 aps,
]{revtex4-2}
\usepackage[dvipsnames]{xcolor}
\usepackage{graphicx}
\usepackage[dvipsnames]{xcolor}
\usepackage{ragged2e}
\usepackage{graphicx}
\usepackage{dcolumn}
\usepackage{bm}

\emergencystretch=\maxdimen
\begin{document}

\preprint{APS/123-QED}

\title{Non-equilibrium phase transitions in coupled nonlinear optical resonators}

\author{Arkadev Roy$^1$}
\author{Rajveer Nehra$^1$}
 \author{Carsten Langrock$^2$}
 \author{Martin Fejer$^2$}
 \author{Alireza Marandi$^1$$^*$}
\affiliation{%
$^1$Department of Electrical Engineering, California Institute of Technology, Pasadena, California 91125, USA}%
\affiliation{%
$^2$Edward L. Ginzton Laboratory, Stanford University, Stanford, California, 94305, USA}
\affiliation{$^*$marandi@caltech.edu}

\begin{abstract}
Phase transitions and the associated symmetry breaking are at the heart of many physical phenomena. Coupled systems with multiple interacting degrees of freedom provide a fertile ground for emergent dynamics that is otherwise inaccessible in their solitary counterparts. Here we show that coupled nonlinear optical resonators can undergo self-organization in their spectrum leading to a first-order phase transition. We experimentally demonstrate such a  spectral phase transition in time-multiplexed coupled optical parametric oscillators. We switch the nature of mutual coupling from dispersive to dissipative and access distinct spectral regimes of the parametric oscillator dimer. We observe abrupt spectral discontinuity at the first-order transition point which can pave the way for the realization of novel transition-edge sensors. Furthermore, we show how non-equilibrium phase transitions can lead to enhanced sensing, where the applied perturbation is not resolvable by the underlying linear system. Our results can pave the way for sensing using nonlinear driven-dissipative systems leading to performance enhancements without sacrificing sensitivity. 
\end{abstract}  

\maketitle

Coupled systems are omnipresent ranging from neuronal connections in biological brains, artificial neural networks, social networks, power grids, circadian rhythms, and reaction-diffusion chemical systems \cite{watts1998collective}. The nonlinear dynamics and the ensuing collective behaviors of coupled systems are remarkably richer than isolated individual ones \cite{tikan2021emergent, grigoriev2011resonant, zhang2019electronically, zhang2021squeezed, miller2015tunable, xue2019super, roy2021nondissipative}. These networks are endowed with complex physics that can have profound consequences in sensing \cite{guo2020distributed} and computing \cite{mcmahon2016fully, marandi2014network}. 

Emergent phenomena in complex systems are ubiquitous and some paradigmatic examples of these non-equilibrium phenomena includes synchronization \cite{jang2018synchronization, fruchart2021non}, and pattern formation \cite{roy2021spectral, haken1975cooperative,vaupel1999observation, cross1993pattern, ropp2018dissipative, taranenko1998pattern, oppo2013self}. Gain competition/ energy exchange among the components of a many-body system in the microscopic scale can lead to emergent macroscopic behaviors \cite{wu2019thermodynamic} including the appearance of Turing patterns \cite{turing1990chemical}, coherent oscillation \cite{degiorgio1970analogy}, and mode-locking \cite{wright2017spatiotemporal, gordon2002phase}. Understanding and engineering phase transitions in driven-dissipative systems constitutes a new frontier of many-body physics and non-equilibrium dynamics \cite{stanley1971phase, prigogine1968symmetry}. Non-equilibrium driven-dissipative systems open new possibilities and opportunities that are not present in their equilibrium counterparts. For instance, time crystal is a non-equilibrium phase of matter that is believed to be realizable in out-of-equilibrium settings \cite{wilczek2012quantum, else2016floquet}. Photonics provides a congenial platform to engineer the drive and the dissipation for the exploration of non-equilibrium emergent phases and dynamical phase transitions \cite{roy2021spectral, tikan2021emergent, dechoum2016critical, drummond1980non, kuznetsov1991optical}.    

Phase transition is associated with the qualitative change in the system behavior as a control parameter is varied across a critical/transition point. An order parameter is often used to characterize systems exhibiting critical behaviors. Discontinuity in the order parameter (its derivative) is a universal signature of first-order (second-order) phase transitions \cite{stanley1971phase}. Such abrupt discontinuities have been leveraged in transition-edge sensors to perform ultra-sensitive measurements down to single-photon levels \cite{gol2001picosecond}. Engineering such discontinuities in driven-dissipative systems is highly desirable to develop high-sensitivity transition-edge sensors that are governed by non-equilibrium dynamics and thereby are not impaired by the slow dynamics that limit their counterparts based on thermodynamic equilibrium phase transitions \cite{yang2019quantum}. A promising approach to quantum sensing involves the exploitation of quantum fluctuations in the vicinity of a critical point to improve the measurement precision. Theoretical studies indicate that sensors based on driven-dissipative phase transitions in parametric nonlinear resonators can be a useful resource in this regard \cite{di2021critical}.

Nonlinearity can potentially endow superior sensing capabilities that can attain orders of magnitude performance enhancement over those that rely on linear dynamics alone \cite{del2017symmetry, wang2015nonlinear, kaplan1981enhancement}. For instance, nonlinearity-induced non-reciprocity can amplify the Sagnac effect in the vicinity of a symmetry breaking instability \cite{kaplan1981enhancement}. Similarly, it has been proposed that operating close to the region of bistability can lead to strong enhancement to refractive-index sensitivity \cite{wang2014nonlinearly}. However, experimental demonstrations of the aforementioned nonlinear advantage remain scarce.  

\begin{figure*}[!ht]
\centering
\includegraphics[width=1\textwidth]{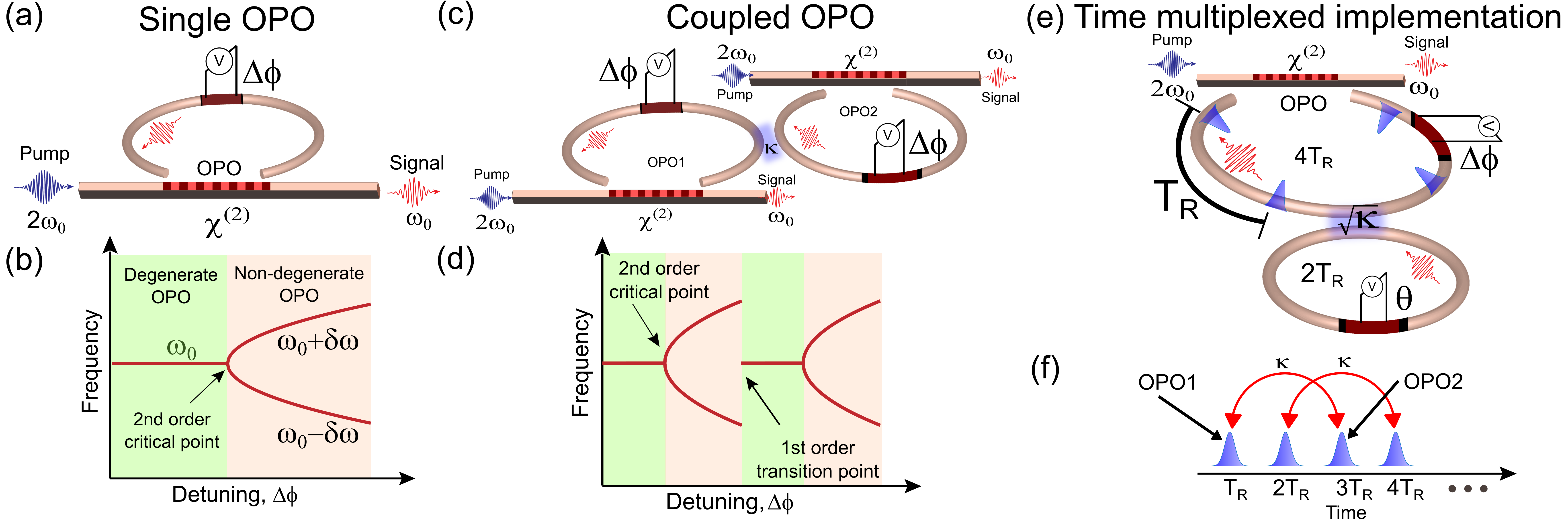}
\caption{\label{fig: schematic} \textbf{Non-equilibrium phase transitions in single and coupled OPOs. } a) Schematic of a single OPO showing the non-resonant pump ($2\omega$) and the resonant signal/idler ($\omega$) interacting via phase-matched quadratic ($\chi^{(2)}$) nonlinearity alongside the detuning ($\Delta\phi$) element. b) Existence of a second-order spectral phase transition in a single OPO where at the critical detuning the OPO transits between the degenerate and the non-degenerate oscillation regimes. c) Schematic of a coupled OPO system with the mutual coupling $\kappa$. d) Existence of a first-order spectral phase transition in coupled OPOs featuring an abrupt spectral discontinuity at the first-order transition point. e) Time-multiplexed implementation of the coupled OPOs consisting of a main OPO cavity (with a roundtrip time of $4T_{R}$) that is twice as long as the linear coupling cavity. The cavity detuning is controlled using a detuning element ($\Delta\phi$) in the main cavity, while the detuning element in the coupling cavity affects the coupling phase $\theta$. f) Illustration of the pulses circulating in the time-multiplexed implementation where the pulse-to-pulse separation is given by the repetition period of the driving pump laser, and the coupling exists between alternate pulses, thereby constituting a coupled OPO system.   } 
\end{figure*}

In this work, we exploit the rich dynamics of coupled optical parametric oscillators (OPOs) to realize non-equilibrium phase transitions. We demonstrate first-order spectral phase transition, and observe abrupt discontinuity at the transition point corresponding to the system's sudden self-organization between degenerate and non-degenerate oscillation regimes. We show that the system of coupled OPOs exhibits qualitatively different behavior with the alteration of their mutual coupling from dispersive to dissipative. We also present nonlinearly enhanced sensing in the driven-dissipative system under consideration where the applied perturbation remains unresolved by the underlying linear system. Our results on non-equilibrium behavior in a system of coupled nonlinear resonators can have far-reaching consequences in the domains of sensing and computing.

\section{Results}

The building block of our coupled system is a doubly resonant OPO which is parametrically driven by a pulsed pump centered around $2\omega_{0}$ (see Fig. 1(a)) \cite{hamerly2016reduced}. The cavity hosts multiple longitudinal frequency modes around the half-harmonic frequency ($\omega_{0}$), where the signal/ idler resides. The distribution of these frequency modes is determined by the cavity group velocity dispersion (GVD, $\beta_{2}$), while the interaction between them is facilitated by the quadratic nonlinearity ($\chi^{(2)}$). The energy exchange between the pump and the signal ($\omega_{0}+\delta\omega$) and the idler ($\omega_{0}-\delta\omega$) modes is governed by the energy and momentum conservation relations. The OPO exists in a trivial state (zero mean-field) below the threshold, which loses stability leading to parametric oscillation as the gain is increased above the oscillation threshold. The oscillation proceeds via the modulational instability, and the OPO assumes a temporal frequency (fast-time scale dynamics) ($\Omega=\delta\omega$, centered around the half-harmonic), corresponding to the maximum growth-rate of perturbations (see Supplementary section 9). $\delta\omega=0$ corresponds to the degenerate oscillation, while $\delta\omega \neq 0$ corresponds to the non-degenerate oscillation regime. The temporal mode with the zero effective detuning experiences the maximum parametric gain. This can happen even in the presence of non-zero cavity detuning, where the GVD-induced detuning counterbalances the linear cavity detuning $\Delta\phi$. This mutual interplay of cavity detuning and GVD leads to a second-order spectral phase-transition as shown in Fig. 1(b) \cite{roy2021spectral}. The critical detuning ($\Delta\phi=0$) marks a soft transition between the degenerate and the non-degenerate parametric oscillation regimes.

\begin{figure*}[!ht]
\centering
\includegraphics[width=1\textwidth]{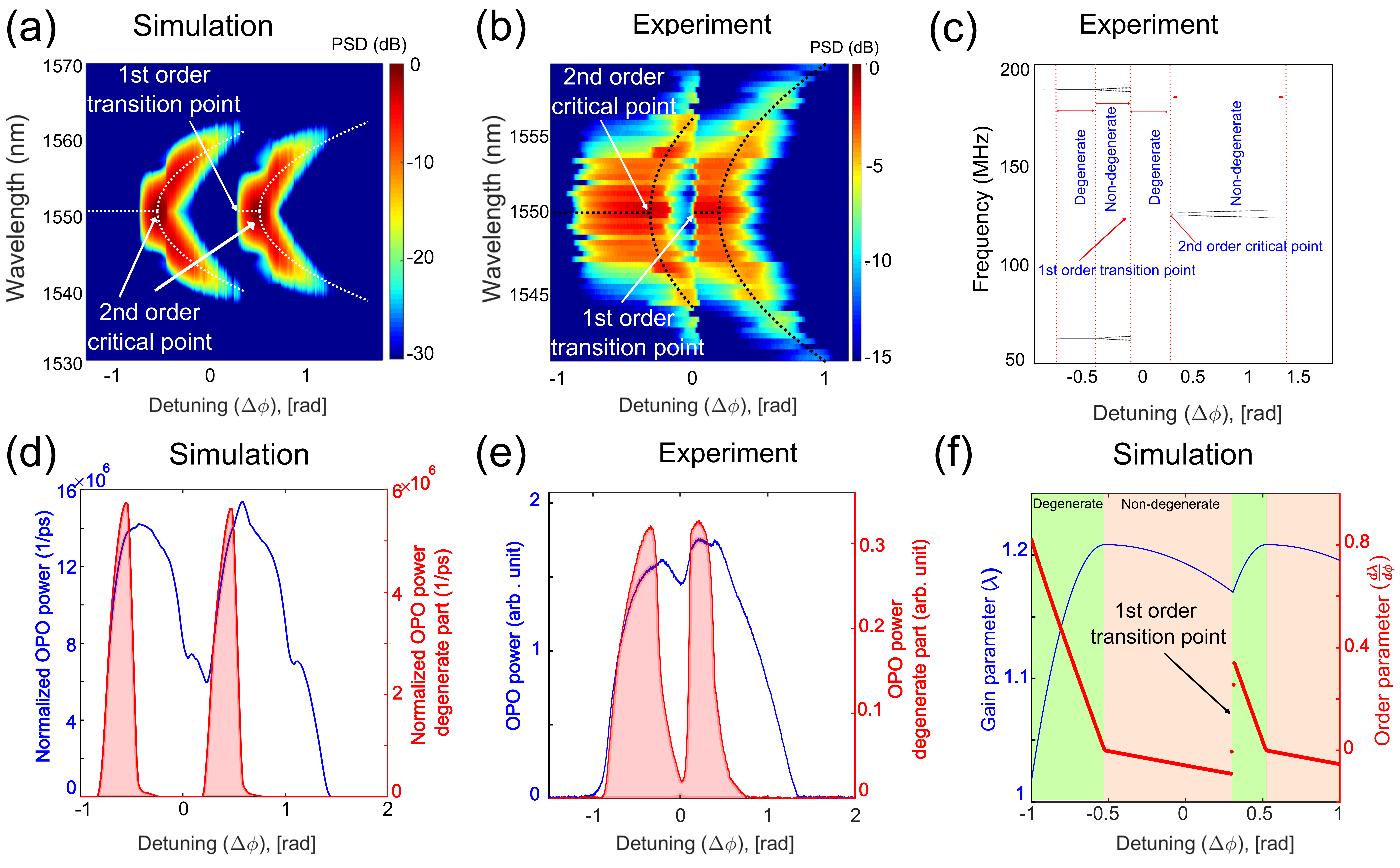}
\caption{\label{fig: Fig2} \textbf{First-order spectral phase transition in coupled OPOs}. a) Numerical simulation of the optical spectrum of coupled OPOs as a function of cavity detuning featuring the second-order phase transitions at the mode-splitting locations as well as the first-order phase transition. b) Experimentally obtained optical spectrum as a function of cavity detuning highlighting the abrupt spectral discontinuity at the first-order transition point. c) Radio frequency beat-note spectrum indicating the distinct degenerate and non-degenerate oscillation regimes demarcated by the second-order critical points and the first-order transition point.  d) Numerical simulation of the OPO power as a function of the detuning. The power contained in the degenerate part of the spectrum (1 nm of bandwidth around the half-harmonic frequency $\omega$) is also plotted alongside showing two distinct degenerate oscillation regimes flanked by non-degenerate oscillation regimes. e) Coupled OPOs power as a function of detuning obtained experimentally. The power contained in the degenerate regime has been extracted using a bandpass filter centered around the half-harmonic frequency. f) Order parameter (derivative of the gain function) shows a discontinuity at the transition point suggesting the existence of a first-order phase transition.     }
\end{figure*}

However, this rich spectral behavior observed in a single OPO doesn't extend linearly with the increase in system size, i.e. to a network of coupled OPOs (see Fig. 1(c)). It is well known that in the realm of parity-time symmetric non-Hermitian systems, increasing the system size, increases the order of the exceptional point \cite{hodaei2017enhanced}. Strikingly, we show that it is possible to realize a hard-transition (first-order transition) in a system of coupled OPOs, where a single OPO is only capable of featuring a soft-transition (second-order transition). Our system of coupled OPOs represents a complex system enabling a rich interplay of nonlinearity, linear coupling ($\kappa$), multimode dynamics, dispersion, drive, and dissipation. This can lead to an abrupt spectral discontinuity between the degenerate and non-degenerate oscillation regimes (see Fig. 1(d)). 

We implement coupled OPOs using time multiplexing \cite{mcmahon2016fully, marandi2014network, leefmans2021topological} (see Fig. 1(e)). This represents a synthetic dimension implementation where the discrete time dimension provided by the equidistant pulses of a mode-locked laser has been utilized to realize a coupled OPO system without increasing the spatial complexity of realizing OPOs in two different cavities. In this two-cavity configuration the main-cavity is twice as long as the coupling cavity. Specifically in our experiments we chose the main cavity round-trip time to be four times the repetition period of the mode-locked laser ($T_{R}$). This ensures that the coupling cavity executes coupling between alternate pulses. Thus pulses occurring at time instants given by $(4n+1)T_{R}$ and $(4n+3)T_{R}$ or $(4n)T_{R}$ and $(4n+2)T_{R}$ (where n is an integer) constitutes two sets of coupled OPOs (see Fig. 1(f)) (see Supplementary section 12). Moreover, our time-multiplexed implementation allows us to mimic different types of coupling (dispersive, dissipative, or hybrid) \cite{ding2019dispersive}, because the phase of the coupling path can be altered by modifying the detuning of the coupling cavity. The detuning elements in the main cavity and the coupling cavity control the cavity detuning parameter ($\Delta\phi$) and the coupling phase $\theta$ independently.  

\begin{figure*}[!ht]
\centering
\includegraphics[width=1\textwidth]{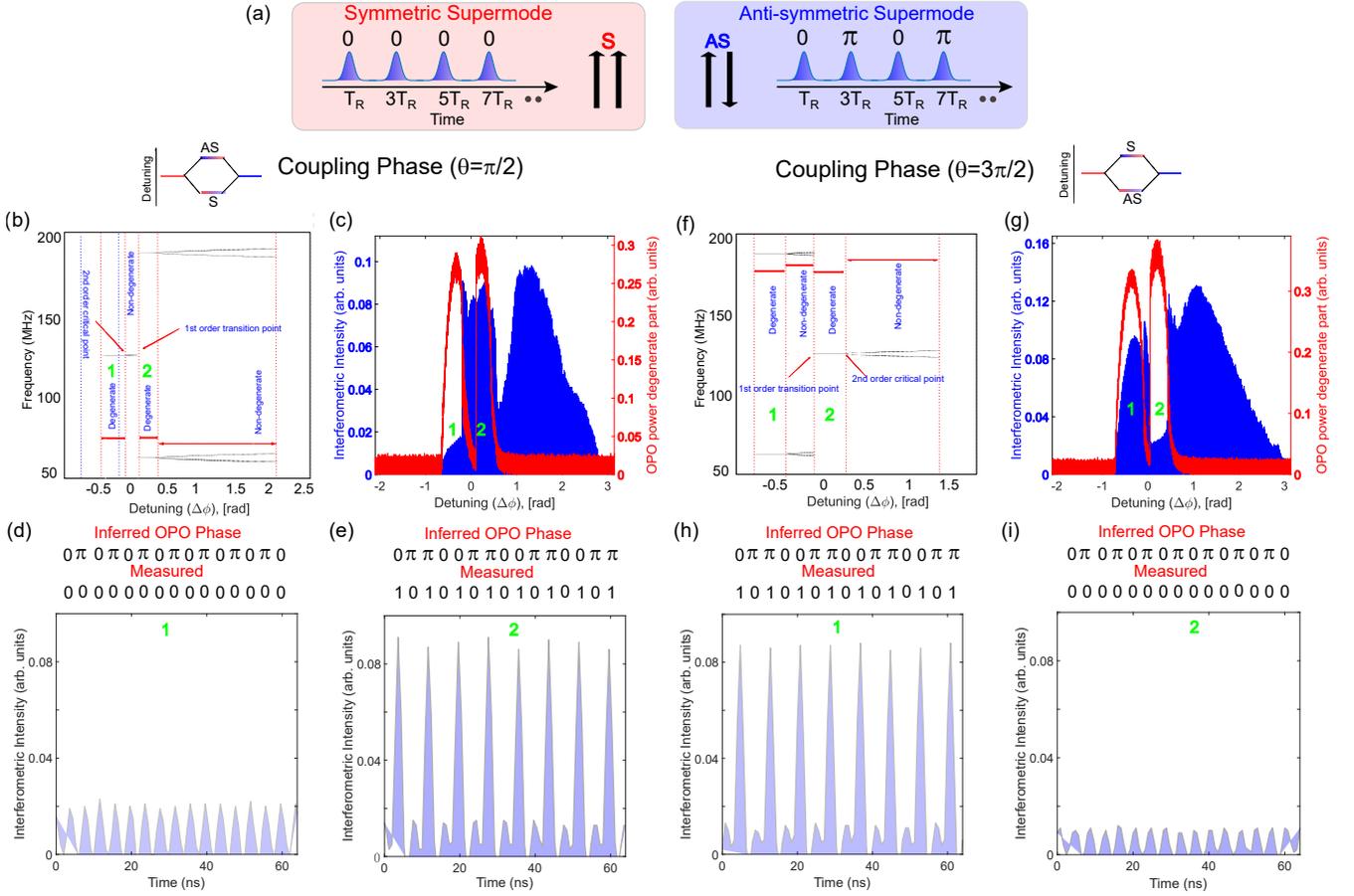}
\justifying
\caption{\label{fig: Fig3}  \textbf{Supermodes of the coupled OPOs.}  a) Illustration of the supermodes and their associated manifestation in the time-multiplexed implementation. b) Radio-frequency beat-note spectrum as a function of detuning in the presence of dispersive coupling with coupling phase $\theta=\pi/2$.  c) Corresponding pulse pattern at the output of a single-pulse delayed Mach-Zehnder interferometer. The OPO power in the degenerate band is also plotted alongside. d) Interferometric pulse pattern in the degenerate regime (marked as 1) shows that the OPO pulses are in-phase representing the symmetric super-mode. e) Pulse pattern in the degenerate regime (marked as 2) shows that the OPO pulses are out of phase representing the anti-symmetric super-mode. Similarly the case with the dispersive coupling and coupling phase $\theta=3\pi/2$ is considered, where the radio-frequency spectrum and the interferometer pulse pattern are displayed in f) and g) respectively. h) The degenerate regime (marked as 1) shows the OPO pulses constituting the coupled OPO are out-of phase implying the anti-symmetric super-mode, while i) shows the degenerate regime (marked as 2) with OPO pulses out-of-phase indicating the symmetric super-mode. }
\end{figure*}

The first-order phase transition in coupled OPOs emerges from the interplay of the supermodes of the coupled cavities and parametric gain. The dispersive coupling $\kappa$ leads to mode-hybridization between the modes of the coupled cavities. These supermodes can be either symmetric when the resonant fields are in phase, or anti-symmetric when they are out-of-phase. The frequency separation between them depends on the coupling strength $\kappa$. At a given excitation frequency, there exists a range of cavity detunings where one of the supermodes is close to resonance, while the other one is off-resonant. In those circumstances, we can consider the dominant supermode only, and the dynamics of the coupled system resembles a single OPO, albeit now in the supermode basis. This results in second-order phase transitions around the mode-splitting points as shown in Fig. 2(a). However, in the range of cavity detunings where the contribution from the supermodes are comparable, there occurs a competition between the two second-order spectral phase transitions (one centered around the symmetric supermode and the other centered around the asymmetric supermode). This gain competition enforces a spectral self-organization of the coupled OPOs leading to a sharp transition between non-degenerate and degenerate oscillation regimes as shown in Fig. 2(a). This proceeds via a first-order phase transition, when the gain of the non-degenerate branch of the symmetric super-mode ceases to be greater than the gain experienced by the degenerate branch of the asymmetric supermode (see Supplementary section 2,3). The experimental results (Fig. 2(b)) of the optical spectrum corroborate the theory and the numerical simulations (coupling factor realized experimentally is lower than the value assumed in the simulation). 

The non-equilibrium phase transitions in coupled OPOs are further characterized by the radio-frequency (RF) measurements (see fig. 2(c)). A sync-pumped doubly-resonant OPO in the non-degenerate regime generates a signal and an idler frequency comb with two carrier-envelope offset frequencies which can be measured through beating with a local oscillator. The abrupt spectral discontinuity of this beat-note measurement unequivocally confirms the occurrence of the first-order phase transition. The output power of the coupled OPOs as a function of detuning is representative of the parametric gain, and leads to maximum conversion efficiencies at the second-order critical points where the supermodes are resonant. This can be seen from the simulation and experimental results in Fig. 2(d,e). The power contained in the spectrum centered around degeneracy is indicative of the degenerate regime of operation. The OPO output after passing through a band-pass filter centered around the half-harmonic frequency is also shown in Fig. 2(d,e) which indicates the presence of two distinct degenerate regimes of operation separated by the non-degenerate oscillation regime. The order parameter (defined as the derivative of the gain parameter with respect to the detuning) exhibits behavior typical of a first-order phase transition with the characteristic discontinuity at the first-order transition point (see Fig. 2(f)). The gain parameter ($\lambda$) reveals the underlying gain competition between the two supermodes.

\begin{figure}[!ht]
\centering
\includegraphics[width=0.5\textwidth]{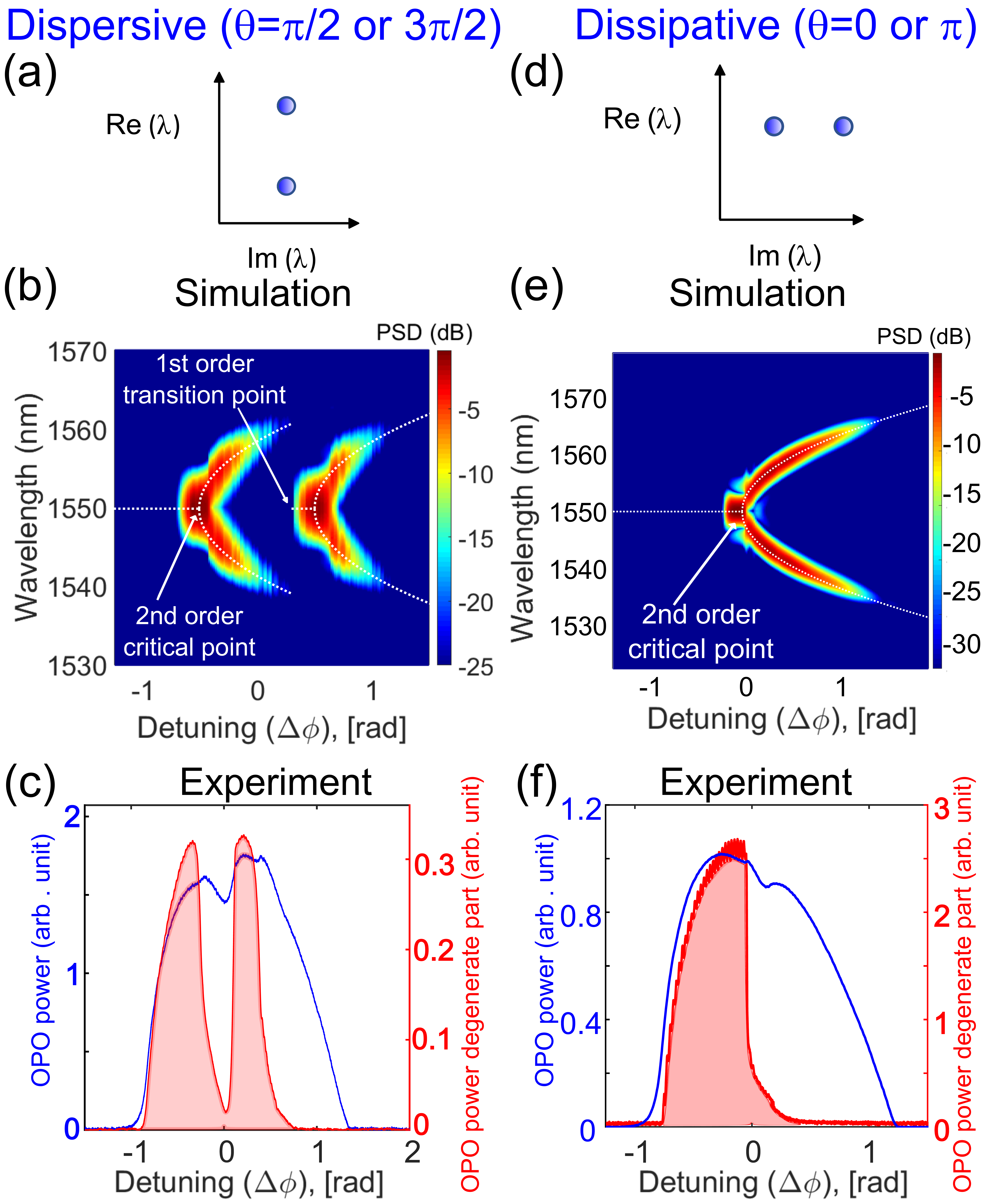}
\justifying
\caption{\label{fig: Fig4}  \textbf{Dispersive vs Dissipative coupling.}  a) Dispersive coupling between coupled resonators results in the splitting of the real part of the eigenvalue (spectrum), where both the super-modes experience identical dissipation. b) The existence of the mode-hybridization in the real part of the spectrum enables the observation of two second-order spectral phase transitions centered around the mode-splitting and the associated first-order transition point. c) OPO power as a function of detuning along with the power contained in the degenerate regime corroborates the existence of the underlying mode-splitting. d) In the presence of dissipative coupling the super-modes experiences different dissipation while their real part remains identical. e) The absence of mode-hybridization in the real part of the spectrum precludes the observation of the first-order phase transition and only leads to features representative of the spectral behavior of a single OPO. f) OPO power as a function of detuning showing the existence of a single degenerate regime confirming the absence of mode-splitting with dissipative coupling.    }
\end{figure}  

The eigenvector composition of the supermodes can be unveiled from the pulse-pattern measurements in the time domain as illustrated in Fig. 3(a). When the coupling phase ($\theta$) equals $\pi/2$, the anti-symmetric eigenmode has a higher frequency (corresponding to larger detuning) than its symmetric counterpart (see Supplementary section 2). The symmetric and the anti-symmetric supermodes have distinct carrier envelope offset frequencies as evident from the RF spectrum (see Fig. 3(b)). The pulse pattern is measured using a one-pulse delayed Mach-Zehnder interferometer (see Fig. 3(c)) to infer the phases of the OPO pulses constituting the coupled OPOs. The coupled OPOs in the symmetric supermode dominated degenerate regime (1) features OPO pulses which are in phase (see Fig. 3(d)). In the anti-symmetric supermode degenerate regime (2), the OPO pulses comprising the coupled OPOs are out of phase (see Fig. 3(e)). When the coupling phase is $3\pi/2$, the frequency spectrum of the supermodes are reversed, with the symmetric supermode now associated with larger detuning. This is revealed in the corresponding measurements shown in Fig. 3(f,g,h,i). This agrees with the results obtained from numerical simulation (see Supplementary section 2).

The spectral behavior of the coupled OPOs drastically differs with the alteration of the nature of mutual coupling ($\kappa$). Modification of the coupling phase ($\theta$) enables us to mimic dispersive ($\pi/2$ or $3\pi/2$), dissipative ($0$ or $\pi$), or hybrid (intermediate phases) (see supplementary section 4,12). Dispersive coupling results in splitting in the real part of the eigenvalues (i.e. mode-splitting) where the supermodes experiences identical rate of dissipation (imaginary part of the eigenvalue is the same) (see Fig. 41(a)). This leads to spectral and temporal features resembling the aforementioned discussions (see Fig. 4(b,c)).  In stark contrast, dissipative coupling leads to splitting in the imaginary part of the eigenvalue where the supermodes experience disparate dissipation \cite{ding2019dispersive}. This property of the dissipative coupling is at the heart of the operation of optical coherent Ising machines \cite{marandi2014network}, and recent demonstrations of topological dissipation \cite{leefmans2021topological}. Consequently, the absence of mode-splitting is also reflected in the spectral (Fig. 4(e)) and the power (Fig. 4(f)) characteristics of dissipatively coupled OPOs. Dissipative coupling precludes the occurrence of a first-order spectral phase transition, and shows the mere presence of a second-order phase transition. 

\begin{figure*}
\centering
\includegraphics[width=1\textwidth]{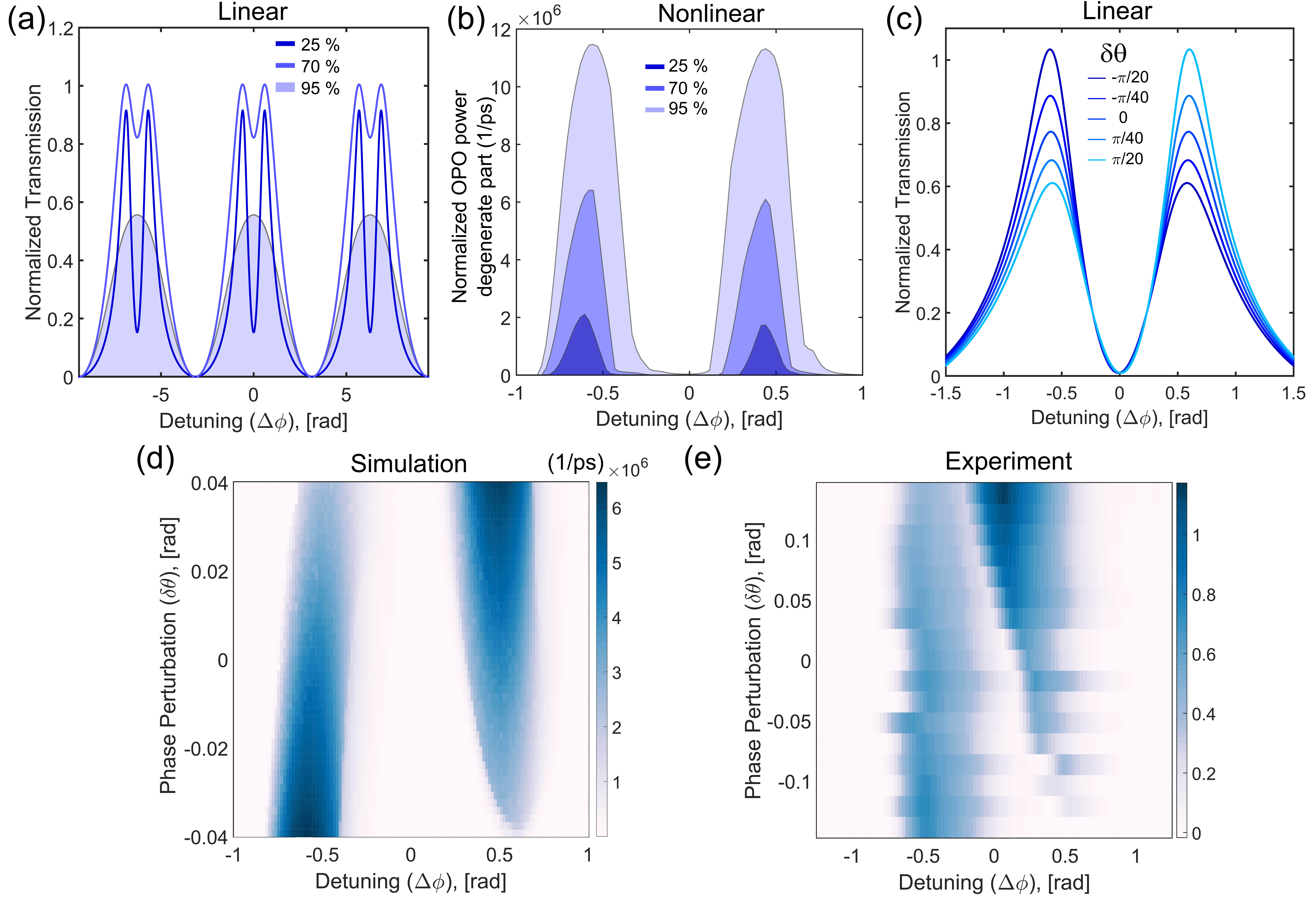}
\justifying
\caption{\label{fig: Fig5}  \textbf{Enhanced sensing using non-equilibrium phase transitions.}  a) Transmission through coupled linear resonators with varying round-trip loss. With lower loss values (high-Q) the mode-splitting is conspicuous, which disappears in the low-Q regime. b) Power contained in the degenerate regime of the coupled OPO as a function of detuning for various values of round-trip loss. In contrast to the linear case, this nonlinear regime could resolve the mode-splitting even in the low-Q regime. c) Transmission in coupled linear resonators (high-Q regime) in the presence of applied perturbation expressed as the perturbation in the coupling phase ($\delta\theta$). The perturbation renders the coupling to be a hybrid of dispersive and dissipative leading to asymmetric mode-splitting. Power contained in the degenerate regime of the coupled OPO with varying perturbation in the low-Q regime obtained through d) numerical simulation, e) experiments. Despite operating in the low-Q regime  the nonlinear dynamics could resolve the underlying asymmetric mode-splitting in response to small coupling-phase perturbation.   }
\end{figure*}  

The presence of non-equilibrium phase transitions with characteristic discontinuities opens up new opportunities in the domain of sensing. High quality factor optical resonators have been utilized for highly sensitive refractive index perturbation measurements \cite{krioukov2002sensor}. However, the requirement of high quality factor for enhanced sensitivity results in an unavoidable trade-off with the bandwidth, and hence limits the sensing speed. The non-equilibrium phase transition in coupled OPOs can circumvent this trade-off. Figure 5.(a) shows the transmission of coupled optical resonators with different round-trip losses. The mode-splitting is observed in the regime of high-Q (lower round-trip loss), while low-Q (high-bandwidth) resonators cannot resolve the mode-splitting structure. Remarkably, this mode-splitting can be revealed even in the low-Q regime in the nonlinear case where the coupled resonators are parametrically driven as coupled OPOs (see Fig. 5(b)). The sensing parameter can be the phase detuning in the coupling cavity of the time-multiplexed architecture leading to perturbation ($\delta\theta$) in the coupling phase $(\theta)$. In the presence of this coupling phase perturbation, coupled high-Q linear resonators will respond with asymmetric mode-splitting, where the degree of the asymmetry depends on the strength of the perturbation (see Fig. 5(c)). The asymmetry also reflects the sign of the phase perturbation which is an added advantage over high-Q linear cavity based simple sensing arrangements where the sensors can suffer from directional ambiguity \cite{heideman1999remote}. This asymmetric mode-splitting behavior cannot be resolved by low-Q coupled linear resonators. However, low-Q coupled OPOs can extract these features which is shown in Fig. 5(d,e) by displaying the power contained in the degenerate part (using a band-pass filter). Results obtained from our low-Q (high gain and bandwidth) experimental setup agrees well with the simulation (see Supplementary section 1). These results indicate the potentials of non-equilibrium phase transitions for enhanced sensing, where the linear counterpart is incapable of resolving the phase perturbation.  

\section{Discussion}

In summary we have demonstrated the occurrence of first-order non-equilibrium phase transitions in coupled OPOs. We experimentally characterized the spectral and the temporal features associated with this phase transition, and also revealed the eigenvector composition of the supermodes. We highlight the distinct spectral behaviors of coupled OPOs in the presence of dispersive and dissipative couplings.  Finally, we demonstrate the potential of these phase transitions in coupled OPOs for enhanced sensing. The results presented here can be directly relevant to other systems including Faraday waves in hydrodynamics and parametrically forced mechanical or chemical systems. With the recent progress in the nanophotonic lithium niobate platform \cite{jankowski2020ultrabroadband, lu2021ultralow,guo2021femtojoule}, exploration of extended lattices can become feasible, paving the way towards the study of emergent nonlinear phenomena in soliton networks and higher dimensional lattices (see Supplementary section 8) \cite{tusnin2021coherent}. The demonstrated phase transition can be modeled using the universal coupled Swift-Hohenberg equation and can be implemented in Kerr nonlinear resonators as well \cite{longhi1996swift, okawachi2015dual}. The abrupt discontinuity at the first-order transition point and the associated spectral bi-stability can open new possibilities in the domain of precision sensing (see Supplementary section 6). The semi-classical regime considered in this work can be probed below the oscillation threshold \cite{wu1986generation}, where a quantum image of the above threshold spectral phase transition exists, which may lead to the co-existence of a quantum phase transition (see Supplementary section 7) \cite{gatti1995quantum}. Our study mainly focuses on the adiabatic regime where the control parameter is varied gradually. Introduction of non-adiabaticity can lead to the Floquet dynamics with enriched phase diagrams (see Supplementary section 13) \cite{longhi2000nonadiabatic}. Intriguing dynamics is also expected in the case of nonlinearly coupled resonators \cite{menotti2019nonlinear}. Analysis of noise mechanisms that could possibly constrain the achievable precision will be the subject of future work. Our work lays the foundation for the exploration of emergent dynamics and critical phenomenon beyond the single-particle description and insinuates potential advances in sensing and computing. 

\section{Methods}
\subsection{Experimental Setup}
 The simplified experimental schematic is shown in Fig 1.e, a detailed version of which is presented as Fig S.1 (Supplemental Section 1). The OPO pump is derived from a mode-locked laser (1550 nm @ 250 MHz) through second-harmonic generation (SHG) in a bulk periodically poled lithium niobate (PPLN) crystal.  The main cavity repetition rate is a quarter of the repetition rate of the mode-locked laser and is composed of a PPLN waveguide (a reverse proton exchange waveguide that phase-matches the interaction between pump photons at 775 nm and signal/idler photons around 1550 nm) \cite{langrock2007fiber} with fiber-coupled output ports, a piezo-mounted translation stage (to provide the cavity detuning), a free-space section (to match the pump repetition rate to be multiple of the free spectral range of the cavity), an additional fiber segment to engineer the cavity dispersion ( dispersion compensating fibers), and a couple of beam splitters (which provide the output coupling and interface with the coupling cavity). The coupling cavity repetition rate is half the repetition rate of the mode-locked laser and comprises of several gold mirrors and a piezo translation stage (to adjust the coupling phase). Additional details pertaining to the experimental setup/methods are provided in the supplementary information (Supplemental Section 1).   \
\subsection{System Modeling}
 The parametric nonlinear interaction in the PPLN waveguide (length $L$) is governed by: 
\begin{subequations}
\label{eq:whole2}
\begin{equation}
\frac{\partial a}{\partial z} =\left[ - \frac{\alpha^{(a)}}{2}  -\textrm{i}\frac{\beta_{2}^{(a)}}{2!}\frac{\partial^{2}}{\partial t^{2}}+ \ldots\right]a +\epsilon a^{*}b
\label{subeq:3}
\end{equation}
\begin{equation}
\frac{\partial b}{\partial z}=\left[- \frac{\alpha^{(b)}}{2} - u\frac{\partial}{\partial t} -\textrm{i}\frac{\beta_{2}^{(b)}}{2!}\frac{\partial^{2}}{\partial t^{2}} + \ldots \right]b -\frac{\epsilon a^{2}}{2}
\label{subeq:4}
\end{equation}
\end{subequations}

 The evolution of the signal($a$) and  the pump($b$) envelopes in the slowly varying envelope approximation are dictated by (1a) and (1b) respectively \cite{hamerly2016reduced}. Here $u$ represents the walk-off parameter, $\alpha$ denotes the attenuation coefficients, and the group velocity dispersion coefficients are denoted by $\beta$.  The effective second-order nonlinear coefficient ($\epsilon$) is related to the SHG efficiency \cite{hamerly2016reduced}. Additionally, the OPO fields experience the effect of the coupling and the cavity feedback every roundtrip. 
  \begin{subequations}
\label{eq:whole3}
\begin{equation}
a_{1}^{'} =\sqrt{1-|\kappa|^{2}}a_{1}+|\kappa|\textrm{e}^{\textrm{i}\theta}a_{2} 
\label{subeq:5}
\end{equation} 
\begin{equation}
a_{2}^{'} =\sqrt{1-|\kappa|^{2}}a_{2}+|\kappa|\textrm{e}^{\textrm{i}\theta}a_{1} 
\label{subeq:6}
\end{equation}
\end{subequations}
where, the subscripts (1,2) refers to the OPO1 and the OPO2 comprising the coupled OPO. The coupling strength is denoted by $|\kappa|$, and $\theta$ is the coupling phase. 

 \begin{subequations}
\label{eq:whole3}
\begin{equation}
a_{(1,2)}^{(n+1)}(0,t) = {\cal F}^{-1}\left\{ G_{0}^{-\frac{1}{2}} e^{i\bar{\phi}} {\cal F} \left\{ a_{(1,2)}^{'(n)}(L,t)\right\}\right\} %
\label{subeq:5}
\end{equation} 
\begin{equation}
\bar{\phi} = \Delta\phi  + \frac{l \lambda^{(a)}}{2c}(\delta\omega)  + \frac{\phi_{2}}{2!}(\delta\omega) ^{2}+ \ldots
\label{subeq:6}
\end{equation}
\end{subequations}

 Eq. (3) includes the round-trip loss which is lumped into an aggregated out-coupling loss factor $G_{0}$, the GVD ($\phi_{2}$) of the cavity and the detuning ($\Delta\phi$) ($\Delta\phi=\pi l$, $l$ is the cavity length detuning in units of signal half-wavelengths) of the circulating signal from the exact synchrony. The round-trip number is denoted by $n$. $\cal F$ denotes the Fourier transform operation. The equations are numerically solved adopting the split-step Fourier algorithm.  \\

\section{Data availability}
The data that support the plots within this paper and other findings of this study are available from the corresponding author upon reasonable request. \\

\section{Code availability}
The codes that support the findings of this study are available from the corresponding author upon reasonable request.

\nocite{*}
\bibliographystyle{unsrt} 
\bibliography{ref}

\section{Acknowledgments}
 The authors thank Dr. Saman Jahani and Dr. Myoung Gyun Suh for helpful discussions. The authors gratefully acknowledge support from ARO grant no. W911NF-18-1-0285, AFOSR award FA9550-20-1-0040, NSF Grant No. 1846273, and 1918549, and NASA. The authors wish to thank NTT Research for their financial and technical support. \\

\section{Author Contribution}
 A.R and A.M conceived the idea. A.R. performed the experiments with help from R.N.   A.R. developed the theory and performed the numerical simulations. C.L. fabricated the PPLN waveguide used in the experiment with supervision of M.F. A.R. and A.M. wrote the manuscript with input from all authors. A.M. supervised the project. \\

\section{Competing Interests}
The authors declare no competing interests. \\


\end{document}